\documentclass[num-refs]{wiley-article}
\usepackage{siunitx}
\usepackage{float}
\usepackage{color}
\usepackage{todonotes}

\papertype{Original Article}
\paperfield{Journal Section}
\title{On the nature of explanations offered by network science: A perspective from and for practicing neuroscientists}

\author[1]{Maxwell A. Bertolero}
\author[1,2,3,4,5,6]{Danielle S. Bassett}
\affil[1]{Department of Bioengineering, School of Engineering \& Applied Science, University of Pennsylvania, PA, 19104 USA}
\affil[2]{Department of Electrical \& Systems Engineering, School of Engineering \& Applied Science, University of Pennsylvania, PA, 19104 USA}
\affil[3]{Department of Psychiatry, Perelman School of Medicine, University of Pennsylvania, Philadelphia, PA, 19104 USA}
\affil[4]{Department of Physics \& Astronomy, College of Arts \& Sciences, University of Pennsylvania, Philadelphia, PA, 19104 USA}
\affil[5]{Department of Neurology, Perelman School of Medicine, University of Pennsylvania, Philadelphia, PA, 19104 USA}
\affil[6]{Santa Fe Institute, Santa Fe, NM, 87501 USA}
\corraddress{Danielle S. Bassett PhD, Department of Bioengineering, School of Engineering \& Applied Science, University of Pennsylvania, PA, 19104 USA}
\corremail{dsb@seas.upenn.edu}

\presentadd{Department of Bioengineering, School of Engineering \& Applied Science, University of Pennsylvania, PA, 19104 USA}
\fundinginfo{MB acknowledges support from the NIH T32 training mechanism (PI: Sheline). DSB acknowledges support from the Alfred P. Sloan Foundation, the Paul Allen Foundation, the John D. and Catherine T. MacArthur Foundation, the Center for Curiosity, and the ISI Foundation.}
\runningauthor{Bertolero \& Bassett}

\begin{document}

\maketitle

\begin{abstract}
Network neuroscience represents the brain as a collection of regions and inter-regional connections. Given its ability to formalize systems-level models, network neuroscience has generated unique explanations of neural function and behavior. The mechanistic status of these explanations and how they can contribute to and fit within the field of neuroscience as a whole has received careful treatment from philosophers. However, these philosophical contributions have not yet reached many neuroscientists. Here we complement formal philosophical efforts by providing an applied perspective from and for neuroscientists. We discuss the mechanistic status of the explanations offered by network neuroscience and how they contribute to, enhance, and interdigitate with other types of explanations in neuroscience. In doing so, we rely on philosophical work concerning the role of causality, scale, and mechanisms in scientific explanations. In particular, we make the distinction between an explanation and the evidence supporting that explanation, and we argue for a scale-free nature of mechanistic explanations. In the course of these discussions, we hope to provide a useful applied framework in which network neuroscience explanations can be exercised across scales and combined with other fields of neuroscience to gain deeper insights into the brain and behavior.
\end{abstract}
\section{Introduction}

In contemporary scientific inquiry both within and beyond neuroscience, the term mechanism is often used when referring to explanations of how the brain works beyond mere description, history, or teleology. We can describe the brain's white matter connections (description), how these connections have changed throughout evolution or morph during development (history), and what these connections exist to do (teleology). But the answers to these questions do not necessarily tell us how white matter works; a mechanistic explanation involves explaining \textit{how} white matter conducts, processes, and sends neural signals across the brain during a particular process. While a mechanistic understanding of white matter involves mere description, history, and teleology, it also goes far beyond them \cite{craver2007explaining,craver2013search}.

Fundamentally, neuroscientists seek mechanistic explanations of how the brain functions to support cognition and behavior. Despite that shared goal, there remains broad disagreement in the field about exactly what types of explanations are mechanistic. Such disagreement tends to hamper cross-disciplinary work, thereby hindering scientific advances. It is therefore timely to consider complementary perspectives. Here we review philosophical work and empirical evidence suggesting that much of the disagreement over the nature of mechanisms in neuroscience could be diffused by (i) separating the notion of ``mechanism'' from that of ``spatial scale'' such that mechanisms can be identified at many different spatial scales, and by (ii) establishing how correlative evidence can support mechanistic explanations. In discussing the former, we summarize a working definition of mechanism that is independent of scale. By scale here, we mean the size of the system's components. In discussing the latter, we review evidence that mechanistic explanations can be used to provide predictions of a system’s structure or function, and we explain how such predictions can be based on either correlative evidence or necessitative evidence (unfortunately often confused with causal evidence). A definition of mechanism that is independent of both scale and the type of evidence will together allow us to link neurons to regions, regions to whole brain dynamics, dynamics to cognition, and cognition to behavior.

In working through these preliminaries, we seek to lay down a foundation for understanding the specific contributions of the emerging field of network neuroscience to the broad and general goals of neuroscientific inquiry. Network neuroscience stems from a thoughtful integration of the mathematics of network science with the biological field of neuroscience in an effort to better understand the physical substrate and consequent function of the mind. The underlying assumptions of the approach are that the brain can be meaningfully separated into units (network nodes) with well-defined interactions (network edges), and that the pattern of inter-unit interactions (network topology) enables the rich complex dynamics observed in the brain to support cognitive function. Although we primarily focus on network neuroscience at the macro-scale where we the authors most frequently contribute, we also consider the instantiation of network neuroscience across a range of spatial scales, and its potential to offer both correlative and causal evidence. Based on this discussion, we consider the types of mechanistic explanations that network neuroscience can offer. Before moving forward, it is critical to note that this work is not a technical or philosophical analysis or a reworking of scale, causality, or mechanism. Instead, we seek to elucidate how the explanations of network neuroscience fit into a more explicit account of neuroscientists' common usage of the terms scale, causality, and mechanism by leveraging work on these topics from the philosophy of science. Here, our main goal is to show how network neuroscience can provide evidence for mechanistic explanations of the brain, going beyond mere description of the brain’s connections and topology, primarily by clarifying the distinction between an explanation and the evidence supporting that explanation.

\section{What is network neuroscience?}

\begin{figure}
\centering
\includegraphics[width=12cm]{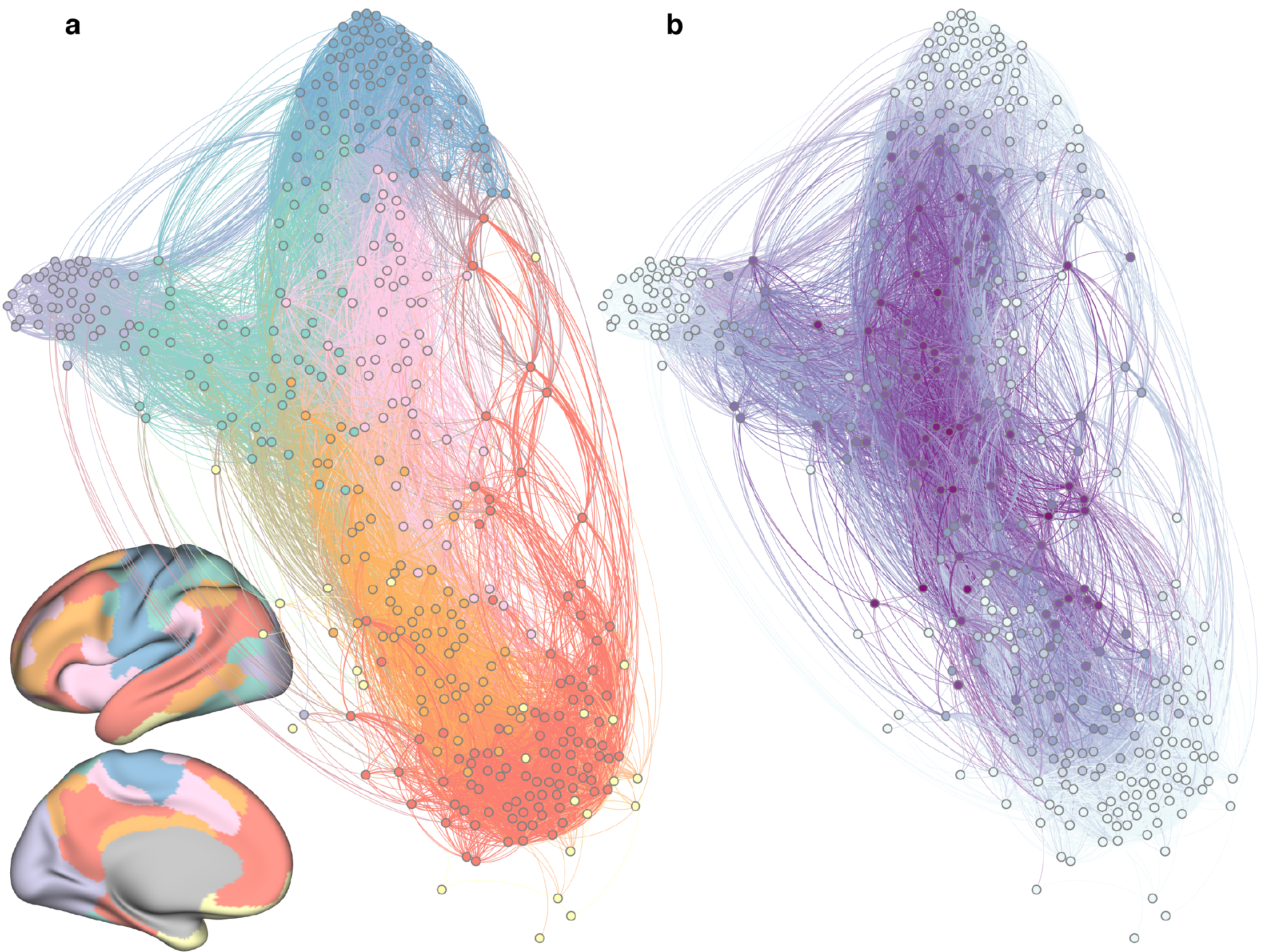}
\caption{\textbf{A network model of functional relationships between brain regions at the large scale in humans.} Each of the 400 brain regions is represented as a network node, which in turn is indicated in this figure by a colored sphere. Each functional relationship between two brain regions is represented as a network edge, which in turn is indicated in this figure by a colored line. \textbf{a}, Here color denotes the assignment of brain regions to putative functional modules that support cognition. Anatomical locations of modules are represented by projecting the color of regions onto the cortical surface of the brain. \textbf{b}, Here color denotes the strength of the participation coefficient, a measure of a node's connectivity to many different modules. Nodes with high participation coefficients are called connector hubs. In both of these layouts, nodes are treated as repelling magnets connected by springs; in this physical representation, nodes that are tightly connected cluster together. Note that connector hubs cluster together at the center of the network, indicative of their role in integration and coordinating brain connectivity across modules.}
\end{figure}

Network neuroscience is an emerging field whose conceptual frameworks, mathematical underpinnings, and applications would require a book \cite{sporns2010networks} or several books \cite{phillips1981fundamentals,bianchi2012network,bassett2017network} to describe comprehensively. Indeed, a full introduction to network neuroscience is beyond the scope here, and has been covered well elsewhere \cite{sporns2010networks}. Here we provide a succinct and non-comprehensive description that will allow a reader to understand the basics of the field and to evaluate our later arguments and examples. Network neuroscience posits that the brain can be usefully represented as a collection of two types of items: (i) nodes, which are typically regions of the brain, groups of neurons, or individual neurons, and (ii) edges, which can either be structural connections, typically in the form of white matter or axons, or statistical dependencies, typically in the form of correlations in regional activity across time \cite{newman2010networks,Bassett:2017e52}. We can decompose this basic network into smaller sub-networks that we call communities or \textit{modules}; each module is composed of nodes that are more tightly interconnected to one another than to nodes in other modules. The division of nodes into modules allows us to measure the role that each node plays in the network topology. One particularly interesting statistic is the participation coefficient, which measures how evenly spread a node's connections are across modules. A node with a high participation coefficient is called a \emph{connector hub}. Specific analyses of the participation coefficient and other network statistics are necessarily descriptive in nature. However, as we will go on to explain, the network model of the brain and how it varies across individuals can be leveraged and combined with theory, computation, and other sources of data, such as genetics, neurology, and behavior, to test mechanistic models of brain function.

\section{What is a mechanistic explanation in neuroscience?}

To contextualize this discussion, we note that the philosophical debate concerning mechanisms is extremely robust. Here we simply summarize a working definition that we view as having a broad consensus between and among philosophers and neuroscientists. We restrict ourselves to features of a mechanistic explanation that are most immediately relevant, but broader and deeper accounts of the topic are well established in the philosophy of science \cite{craver2007explaining}. For further details, we point interested readers to relevant and important debates concerning computational mechanisms \cite{milkowski2013explaining}, the existence of mechanisms in dynamic complex systems \cite{chemero2008after}, and the question of whether mechanisms are necessarily linked to scientific realism \cite{colombomech}.

\subsection{A neuroscientist's working definition}

To construct a mechanistic explanation of a system exhibiting a particular phenomenon, one must decompose that system into its relevant parts and explain how they are organized as well as how they interact to produce the phenomena \cite{craver2007top,colombomech,machamer2000thinking}. A mechanistic model explains a system's phenomena in virtue of its parts, their operations, and their organization, which can together produce the phenomena that is to be explained via a set of orchestrated interactions \cite{bechtel2005explanation}. Such explanations offer a mechanism that must do the work in a causal way \cite{craver_networks,craver2007explaining}, rather than arriving at the state of “work done” via a set of correlative relations or temporal sequence of events. Finally, this explanation must allow for an accurate manipulation of the system \cite{colombomech}. Consider a dirt bike\footnote{We note that the dirt bike analogy differs from the brain in two key aspects: the dirt bike is linear while the brain is nonlinear, and the dirt bike is composed of single-function components while the brain is composed of multi-function components; while these differences are intimately connected with reasons that network tools are useful in neuroscience, we keep the analogy simple to ensure that our basic arguments are accessible to a broad readership. Moreover, one of the authors (we leave it to the reader to deduce which) races dirt bikes, and thus the analogy is particularly \emph{apropos}.}. Force is created by the combustion in the motor and enacted upon the crankshaft, which is connected to a front sprocket; via a chain, the force is transferred onto the rear sprocket, which is connected to a hub, which, via spokes, is connected to a rim, which has a rubber tire mounted in it, which has knobs that grip dirt. In this explanation, we describe the functional role of entities and causal mechanistic relations between them (e.g., force is transferred between the two sprockets via a chain), rather than merely describing physical characteristics of the entities (i.e., the sprockets are toothed aluminum wheels). We know what a dirt bike was built to do (its teleology: traverse dirt) and how it does so. Moreover, we can determine which part is broken based on particular behaviors; if the front sprocket is spinning but not the rear sprocket, the chain is likely broken but the motor is likely intact. 

Neuroscientists in practice tend to adopt these requirements while defining a mechanism. Across spatial scales of inquiry, mechanistic explanations in neuroscience answer the question ``How does the brain work?'' in a similar manner to how we would answer the question ``How does a dirt bike work?''. We will consider two examples: one at the large scale, and one at the small scale. The first example is that of word learning in cognitive neuroscience. Wernicke’s area has been associated with the comprehension of speech, whereas Broca’s area has been associated with the production of speech. Humans require both areas to learn a new word and to employ it, and therefore the physical connection that allows the two regions to communicate, the arcuate fasciculus, is crucial \cite{lopez2013word}. Macaques, who do not have the human capacity for spoken language, have an arcuate fasciculus, but it is not left-lateralized, and it is smaller than it is in humans\cite{eichert2019special}. Damage to this connection is followed by an inability to learn new words, and specific word learning deficits can be traced to damage of Broca's area, Wernike's area, or the arcuate fasciculus, establishing that these areas are \textit{necessary} for the function. Moving beyond necessity to correlative evidence, recent studies have demonstrated that humans with more robust white matter tracts in the arcuate fasciculus exhibit better language learning abilities \cite{wong2011white} and this pathway strengthens during the development of language \cite{broce2015fiber}. Together, these correlative and necessitative results are consistent with (but do not prove) the casual mechanistic explanation that the arcuate fasciculus transfers information during word learning. 

It is important to note that this explanation glosses over, but depends on, smaller scale mechanistic explanations of neural coding and communication. At this smaller scale, our second example is that of navigation. Different types of cells in the medial entorhinal cortex represent different aspects of navigation via the mechanism of feature detection of the sensory cortices. Grid cells respond to the animal’s location in the environment, border cells express the animal’s proximity to geometric borders, speed cells reflect the running speed of the animal, and head direction cells indicate the orientation of the animal relative to landmarks in the environment \cite{rowland2016ten}. The mechanism here is a neural mapping of the animal moving in the world. 

Note that in the two examples of mechanistic explanations we just discussed, there existed a notion of causality, even though necessity, not causality, is observed. Scientists in general, including neuroscientists, typically emphasize causality in mechanistic explanations \cite{Salmon1984,woodward2005making}. Similar to the manner in which a chain does the work of transferring force, the arcuate fasciculus does the work of transmitting information during word learning, and grid cells do the work of encoding location during navigation. Yet despite the fact that notions of causality rightly accompany notions of mechanism, what neuroscientists unfortunately often mean when they say causality is just necessity. If a region of cortex is active during a cognitive process, and damage to that region impairs that cognitive process, we know that that region is necessary for that cognitive process; inaccurately, the region is also sometimes described as the cause of that cognitive process. It is critical to acknowledge that the notion of necessity is independent from the notion of causality, and a necessary component of a process need not be a mechanism. For example, if the only evidence that the arcuate fasciculus transmits information during word learning was that damage to it impairs word learning, we would not have a mechanistic model, just a necessary relationship between the arcuate fasciculus and word learning. 

To have a mechanistic model, we need multiple lines of evidence, both from establishing necessity and finding correlative evidence, as described above, because both lines provide evidence for a causal mechanism, even though neither are identical with it. Returning to the dirt bike, it is one thing to know that the chain is necessary for the rear wheel to spin. But so is the engine, the throttle, \emph{et cetera}. It is critical to know that the speed of the chain correlates with the rotational speed of the rear wheel. We need both to have a casual mechanism. Finally, it is important to realize that ofttimes descriptions are a key component of a mechanism. It is not trivial to know that the front and rear sprockets are connected via a chain, just as it is not trivial to know that Broca's and Wernike's are connected via the arcuate fasciculus.

\subsection{Where our difficulties arise}

Despite our quest for mechanisms, we as neuroscientists do not often employ technical definitions of them. We seem to operate on some common and unspoken knowledge about what constitutes a mechanism. We know one when we see one; or, at least when we want to. However, this approach tends to progenate misunderstanding, bias, and confusion. An important antidote is to appreciate how mechanistic can be defined, and how that definition might be distinct from notions of necessity and the spatial scale at which we each work. Drawing on efforts in the philosophy of science as well as recent advances in network neuroscience, we summarize a notion of mechanism that is supported by both correlative and necessitative evidence and allows us to link work across scales and methods.

\subsubsection{Causality}

We aim to distinguish a mechanistic explanation from the source of evidence for it. To do so, we must first make the distinction between necessity and causality, which is a feature of a mechanistic explanation. Although we do not attempt to define causality precisely here, knowledge of necessity is certainly not knowledge of causality. And even though causality is a required feature of a mechanistic explanation, a mechanistic explanation (or model) can be supported by either correlative or necessitative evidence, or both. In other words, a mechanistic explanation is a model that we posit to explain a system, and then we seek to obtain evidence of various kinds to support that model and to confirm its verity. The distinction between an explanation and the evidence supporting that explanation is well known to philosophers \cite{craver2013search,craver2007top,betchel2008,betchel2008,betchel2012}, but is less broadly appreciated by neuroscientists. Of course, as we outlined above, mechanistic explanations rely on necessary relationships, and necessitative evidence is valuable. However, necessity, on its own, is not causality, and correlative evidence can be just as valuable in supporting a mechanistic model.

In neuroscience, an emphasis on so-called causal evidence has motivated lesion and ablation studies, as well as stimulation and optogenetics studies. While important, such studies are less inherently valuable in and of themselves than they are when performed explicitly to test a mechanistic explanation that has been formally constructed from different types of evidence. For example, consider a thought experiment in which we destroy a particular brain region that functional neuroimaging has implicated in a particular cognitive process. Because the animal would no longer be able to engage in that cognitive process, one might (wrongly) say that we have uncovered evidence that that region causes that function. However, this is where neuroscientists equating necessity with causality can lead to failure; it is entirely possible that that region is in fact upstream of the region actually performing the relevant computation, and thus the lesion study provides some evidence but not sufficient evidence for a causal mechanistic explanation. In the parlance of our dirt bike analogy: if a dirt bike chain breaks and the rear wheel stops turning, we cannot with certainty infer that the chain is \textit{generating} force. Nor should we. One must measure the whole system to prevent inaccurate inferences, and network approaches are one way to do exactly that. 

As an example, consider a model in which connector hubs integrate information and maintain modular processing in the brain. One could perform a between-subjects analysis to demonstrate that, when a network has strong connector hubs, the network is more modular and cognitive performance is higher \cite{Bertolero:2018e52}. Moreover, when connector hubs are damaged, modularity decreases \cite{gratton2012focal} and cognitive deficits are widespread \cite{warren2014network}. Such correlative evidence, particularly when potential confounds are modeled quasi-experimentally \cite{Marinescu:2018e52} and coupled with necessitative studies can strongly serve to support a mechanistic explanation. While lesion analyses demonstrate necessity, network analysis measures the entire system; both can provide evidence in support of a mechanistic model, \textit{particularly} when combined. In the final section, we describe how this can occur in greater detail.

Obtaining correlative evidence for mechanistic explanations remains critical for the continued advancement of science and may play an increasingly important role in neuroscience for two reasons. First, the types of data available have changed fundamentally in their nature. Concerted efforts aligned with federal and international funding priorities have culminated in enormous repositories of brain, behavior, and genetic data from thousands of individuals \cite{HCP,okbay2016genome}. Such data will be invaluable in constructing descriptive explanations, and in providing correlative evidence for mechanistic explanations. Indeed, cognitive scientists now frequently go beyond the analysis of small datasets and well-controlled studies, instead analyzing large and complex observational data \cite{Griffiths}. In meeting the opportunities that these new data bring, it may prove useful to learn from our colleagues in astronomy and astrophysics who generate large scale observations from noisy data viewed from far away, and then use those observations to inform smaller-scale laboratory experiments \cite{Griffiths}. Mechanistic models can be built from the large scale observations and then confirmed in laboratory experiments that exert greater control over the system \cite{craver2016explanatory,zednik2019models}. We envision such integration between large-scale data analysis and small-scale laboratory experiments to become increasingly prevalent and fruitful in neuroscience.

The second reason that correlative evidence for mechanistic explanations may play an increasingly important role in neuroscience is that many phenomena -- across all domains of biology -- appear to be driven by network-level processes \cite{alon2007simplicity,zednik2019models}. Understanding and manipulating causal structures in such networks is an important area of ongoing research. Yet, finding causality in any system is difficult, but defining causality in networks, isolating causal relations in networks, and experimentally testing causal processes via finding necessitative relationships in networks is extremely difficult \cite{noual2016causality}. To offer a bit of intuition, one simple difficulty lies in the question of whether specific edges or sets of edges within the network are the true driving force, or whether the mechanism is in fact an emergent property of the network as a whole. Determining the answer to this question might require a combinatorially large set of experiments, which could be impractical. A second notable difficulty lies in the fact that many networks associated with biological phenomena are not simple tree-like structures, with linear paths along which a causal chain can be identified, but instead contain nontrivial clustering in addition to complex looped structures and cavities \cite{betzel2018diversity,betzel2018specificity,sizemore2018cliques,reimann2017cliques}. While it remains important to posit causal mechanistic models of network interactions, the predictions of those models can be best evaluated in large correlative analyses of expansive data sets; distinct necessitative manipulations can instead by used to probe highly specific and constrained aspects of the network at a single time, informed by the large-scale correlative evidence.

Finally, some have questioned the value of network neuroscience models, and particularly the correlative nature of models that describe the statistical dependency between activity time courses of two regions \cite{craver_networks}. However, it has been well argued that even though functional connectivity is not itself a mechanism, models of functional connectivity can provide evidence for the mechanisms that cause those correlations \cite{zednik2019models}. In other words, network neuroscience models of functional connectivity can provide rough mechanistic approximations of the brain's component parts and interactions at a large scale \cite{zednik2019models}. A network edge defined by a correlation can do causal mechanistic work; and, a causal mechanism can predict the presence of a correlation, which can then be observed in empirical data. Moreover, network neuroscience explanations are most satisfying when they move beyond a static and mere description of the network's composition and organization. Ideally, such models test mechanistic explanations of brain function by also levering simulation and dynamical models \cite{zednik2019models,Bertolero:2015e52,Bertolero:2017e52}, individual differences in network composition \cite{Bertolero:2018e52,shine2019human}, and lesion analyses of the network \cite{gratton2012focal,warren2014network}. 

In summary, mechanistic models posit causal relationships between the organization of the system and the phenomena to be explained. However, causality in the brain is quite opaque, and we typically inaccurately conflate causality and necessity in neuroscience. Moreover, correlative evidence from network models can certainly bear on the validity of a mechanistic explanation that includes causal relationships, despite the fact that the model's organization and interactions can be quantified from correlations. In particular, this approach is fruitful when combined with necessitative analyses. Thus, the network perspective is increasingly critical to explaining brain function, as the global analyses that can leverage large data sets can inform and constrain interpretations of more localized causal manipulations.

\subsubsection{Scale}

When investigating a given system through the lens of science, we often either explicitly or implicitly choose the scale at which we think we can gain a mechanistic understanding. Ca\textsuperscript{2+} ions exist at a scale that might appear to be useful for gaining a mechanistic understanding of how neurons fire and thereby release neurotransmitters \cite{Craver_Brain,katz1968role}. Yet, this scale does not address the molecular composition and function of the active zone of a presynaptic nerve terminal, which allows for the synaptic vesicle exocytosis that occurs when neurotransmitters are released \cite{Sudhof:2012e52,shin2014exocytosis}. Similarly, this scale does not address the cognitive context that can explain why neurons fire in a particular spatiotemporal pattern. In fact, mechanistic explanations exist at each of these scales separately; no scale is privileged in its potential to offer a mechanistic explanation \cite{craver2007explaining}. 

Returning to the dirt bike analogy, force being transferred from the front sprocket to the rear sprocket via the chain is a relatively high level explanation that does not involve the individual links of the chain or the number of teeth on the sprocket, which determine how force is transferred, but it is also a lower level explanation than one addressing how the chassis and engine work together to propel the bike across dirt. Despite differences in scale, all three explanations can be mechanistic explanations. Similarly, an explanation of brain function involving multiple brain regions communicating via white matter tracts and coordinated activity would not \textit{necessarily} be any less mechanistic than an explanation involving multiple cortical neurons communicating via axons and synapses. For example, consider explanations of various features of visual perception. At a microscale, primary visual cortex -- the earliest cortical area associated with the perceptual of visual stimuli -- contains neurons that temporally coordinate their activity patterns to encode the orientation of a stimulus \cite{gray1989stimulus}. At a macro-scale, information travels between the visual cortex and the posterior parietal cortex \cite{Knudsen_annurev,Andersen1997}, the latter mediating selective attention to motion by modulating the effective connectivity from early visual cortex to the motion-sensitive areas in visual cortex \cite{friston2000attentional}. In both cases, the functional mechanism underlying the cognitive process lies in neurons, or groups of neurons, communicating via axons and coordinated activity. 

The key differences are (i) the scale of the explanation, which does not inherently make an explanation more or less mechanistic, and (ii) the specific function that we wish to explain (orientation tuning or motion detection), which can determine the scale of explanation that is most appropriate. While this distinction has been noted by philosophers  \cite{craver2007explaining}, neuroscientists tend to favor the scale of their work as the scale with strongest mechanistic explanations. This bias is in some sense quite rational; neuroscientists should work at the scale they believe is the most fruitful, and a good explanation at one scale need not derive from a good explanation at another scale \cite{craver2007explaining}. However, a key problem with hegemony of a single scale is that good mechanistic explanations in neuroscience can also integrate across all scales, interdigitating data across various methods \cite{craver2007explaining}. Thus, we must be open to explaining the brain at each scale mechanistically, and also deriving explanations of brain mechanisms that bridge phenomena across two or more scales.

The notion that no specific scale of scientific investigation is privileged in terms of its capacity to provide a mechanistic explanation is also broadly understood across a range of disciplines. But perhaps the discipline that most cleanly discusses the notion -- and has the longest history of utilizing it to understand our world -- is the field of physics \cite{machta2013parameter}. General relativity offers fundamental laws that are required to provide mechanistic explanations on the cosmological scale. Newtonian mechanics offers fundamental laws that are required to provide mechanistic explanations on the scale of phenomena observable by the naked human eye. Quantum mechanics offers fundamental laws that are required to provide mechanistic explanations on the atomic scale. But the specific form of the mechanism or explanation important for one scale is irrelevant at other scales. Scales are related to one another and yet mechanistic explanations at one scale can be independent of those at another scale; macroscopic observables at a larger scale show weak dependence on microscopic details at any of the scales below \cite{cardy1996scaling}.

This perspective is particularly useful when we consider the types of mechanistic explanations that we can seek in neuroscience. Reduction and coarse-graining -- which we often use to move up scales from individual cells to brain regions -- do not either increase or decrease our potential to unearth mechanisms. Instead, they extend the spatial or temporal extent over which the mechanistic explanation holds true, even if one does not reduce to the other, similar to Newtonian and Quantum mechanics. Take spatial navigation. As Craver puts it: ``The influx of Ca2+ ions (atoms) through the NMDA receptor (molecules) initiates the sequence of events leading to LTP (cells), which is part of the mechanism for forming a spatial map in the CA1 region (organs). Map formation is part of the explanation for how the mouse (whole organism) navigates through familiar environments (ecosystems) and among conspecifics and predators (societies)'' \cite{craver2007explaining}. The microscale, mesoscale, and macroscale explanations differ in their content and supporting evidence, but all remain mechanistic in their type, despite the fact that they do not easily reduce to one another. Instead, they all constrain the ways in which we think about the mechanisms underlying the behavior.

\section{Network explanations at the largest scale}

At the largest scale, network science models the brain as approximately 100 to 1000 regions that are connected either physically by white matter tracts or statistically by shared information between regional time series \cite{Bassett:2017e52}. It is therefore particularly relevant to consider the question of how such large-scale network models can offer high-level mechanistic explanations of how the brain works. This question has been extensively covered by philosophers \cite{bechtel2017systems,Colombo2018,craver_networks,rathkopf2018network}. Thus, what we seek to do here is to offer a practical perspective, with recent and prominent examples from the field. We will constrain ourselves to two broad questions: (i) why we should view network neuroscience as offering both parsimonious mere descriptions and mechanistic explanations of brain function, and (ii) how can we decipher between the two, given the above definition of mechanism.

A particularly notable strength of large-scale network neuroscience lies in its ability to study every region of the brain in a single cohesive model, providing intuitions for the functions of complete circuits. A disadvantage is that much local information about the processes that occur or the structures that exist within a node are largely hidden. Such internal processes and structures are instead considered by models constructed at lower scales, where -- particularly in non-human species -- one can measure individual neurons in a region, oblate neurons in that region, and genetically modify the organism to alter the structure of that region to probe local functions.

To further appreciate the utility of network science, it is useful to contrast the types of explanations it can offer with the types of explanations offered by other approaches, and to the assess which types of explanations neuroscientists find satisfying. Let us consider cognitive neuroscientists as an example. Typically, they might seek answers to questions such as: how does a brain region (or a set of brain regions) execute a particular cognitive process? For example, how does the hippocampus store and represent spatial information? How does the orbital frontal cortex represent value? Now imagine that -- for every cognitive process -- we have obtained a satisfactory mechanistic explanation. When someone asks us how the brain works, do we simply hand them this list of so-called explanations? Such a list would be a valuable start, but a set of independent mechanistic explanations in different conceptual languages of disjoint processes cannot fully explain how the brain, as a whole, works. Ideally, we wish to have a language in which to comprehend the function of the entire brain, and this is explicitly what network science has the potential to offer.

Before explaining how network neuroscience can provide mechanistic explanations of the brain, it is important to note that network models at the large-scale can offer simplified mere descriptions of the above brain-behavior relationships. A particularly notable simplification is in a study that reports a significant link between the presence of a pattern of whole brain connectivity within each individual to many behavioral (working memory capacity, spatial reasoning) and demographic (education, income, IQ, life-satisfaction) measures in those individuals using canonical correlation analysis \cite{Smith:2015e52}. Measures that were correlated with the presence of the connectivity pattern tended to be positive personal qualities or indicators (e.g., high performance on memory and cognitive tests, life satisfaction, years of education, income). Measures that were anticorrelated with the presence of this pattern tended to be negative personal qualities or indicators (e.g., those related to substance use, rule breaking behavior, anger). This set of findings suggests that there may be a general pattern of healthy brain function associated with a specific pattern of network-level connectivity. In the same vein, network neuroscience models have the ability to reduce the complexity of descriptions of how mental illnesses emanate from the brain, and to discover dimensions of mental illnesses that neither regional studies nor behavioral analyses can uncover \cite{xia2018linked}. Network approaches have proven useful in discovering biotypes that cannot be differentiated solely on the basis of clinical features, but that are associated with differing profiles of clinical symptoms or treatment response \cite{downar2014anhedonia,Drysdale:2016e52}. Here, the strength of a network model lies in the fact that it can describe connectivity patterns that map in a non-trivial but still simple way to all behaviors. Such models provide striking descriptions, but not explanations, of brain function \cite{hommel2019pseudo}. To move from description to explanation requires that the description offer evidence for a mechanistic model; for example, if the model predicted the above correlations, then the correlations would be evidence in favor of the model.

In addition to offering parsimonious descriptions, network models at the large-scale can be used to generate and test macro-level mechanisms of how the brain works. Note here that much of the evidence involves correlations in empirical data or the results of numerical simulations. However, unlike the studies described in the previous sections, what we empirically or \emph{in silico} observe about human brain networks is tested against a mechanistic model, not presented in isolation as a mere description. Consider a candidate mechanistic explanation of global brain function, which posits that some regions are informationally encapsulated while other regions are informationally integrated  \cite{Fodor}. Let us suppose that the function of a given module (\textit{A}) is largely independent of other modules. Then, we would expect to observe that the activity of module \textit{A} would not increase when other modules were active. If instead we were to observe that the activity of module \textit{A} increases in proportion to the number of other modules active, we would conclude that information from these other modules is relevant to module \textit{A}, causing an increase in computational complexity and thus activity. In this case, we would infer that information processing in module \textit{A} is unlikely to be encapsulated \cite{Fodor}. In our model, regions whose activity scales with the number of modules engaged in a task are likely to be executing computations that are more complex, requiring the integration of information across modules or the tuning of connectivity across modules.  

Recently, we explicitly tested this model in empirical fMRI data from 10,000 experiments and 83 different cognitive tasks ranging from simple finger tapping to working memory. A network is constructed in which brain regions are represented as nodes and correlations in regional activity are represented as network edges. Modules are defined as groups of brain regions with dense interconnectivity. We determined how activity within each module varied with the number of modules engaged in each task. We report that modules composed of primary regions implicated in vision, sensation, and motion do not increase in activity in proportion to the number of modules involved across the 83 tasks. In light of our model, this behavior suggests that those modules are informationally encapsulated \cite{Bertolero:2015e52}. In contrast, frontoparietal and attention modules, which is where most connector hubs are located, do increase in activity in proportion to the number of modules involved across the 83 tasks. In light of our model, this behavior suggests that these modules are not informationally encapsulated but instead perform computationally demanding functions when more modules are engaged in a task. The data support the notion that modules with many connector hubs integrate information or tune whole brain connectivity \cite{Bertolero:2015e52}. 

In this example, empirical evidence and a network model are used to test one of the most debated hypotheses in neuroscience and philosophy of mind  \cite{Fodor,ColomboMod,bertolero2014holism}. The network represents correlations in regional activity. Moreover, the mechanistic model makes a correlative prediction: that the level of activity in frontoparietal and attention modules is positively correlated with the number of modules engaged in the task, whereas the level of activity in sensorimotor modules is not correlated with the number of modules engaged in the task. Despite the fact that both data and model involve correlations, the explanation of how the network functions is mechanistic, with connector hubs doing the mechanistic work of integrating information and tuning whole brain connectivity, which allows other modules to remain relatively independent.

A potential mechanistic explanation that is more local but still quite global seeks to address the function of the unencapsulated connector hubs. Connections from such regions are relatively evenly spread across all modules, making them ideally located to tune connectivity between and among other modules \cite{Guimera:2006e52}. In a series of cross-subject analyses, including a quasi-experimental structural equation model \cite{Marinescu:2018e52}, a recent study we conducted offered evidence that these nodes do indeed tune (borrowing the term from its common use at the neuronal scale \cite{sakai1994neuronal}) the connectivity of other networks, thereby maintaining the network's modular structure \cite{Bertolero:2018e52}. Critically, the more connector hubs were able to tune the network to be modular, the better the subject performed on a range of 50 distinct cognitively demanding tasks. We then gathered merely descriptive experimental evidence suggesting that connector hubs are densely interconnected to each other, forming a \emph{diverse club}\cite{Bertolero:2017e52}. Moreover, when connector hubs are damaged, modularity decreases \cite{gratton2012focal} and cognitive deficits are widespread \cite{warren2014network}. Then, in a series of numerical experiments, we simulated evolutionary algorithms to obtain evidence that this club is only naturally selected if the cost function balances modularity and efficient integration \cite{Bertolero:2017e52}. This result evidences the previously discussed mechanistic explanation that these connector hub nodes coordinate connectivity between modules to maintain the modular structure of the brain while also supporting integration. Note here that machine learning in the form of a deep neural network was used to relate connector hub function to cognitive performance across individuals. But, such machine learning algorithms do not constitute mechanisms on their own; to reach towards mechanism, we must posit and test a mechanistic model. This work posited a mechanistic model that predicted the ability of connector hub function to predict cognition, which was confirmed via machine learning. In sum, we gathered correlative, necessitative, and description evidence to support a mechanistic model.

The tuning function of connector hubs can be contextualized as a network science language explanation of known mechanisms of cognitive control, which is a capacity observed mostly in frontoparietal connector hubs to exert top-down influence over other areas of cortex. Recent evidence supports this putative mechanism by demonstrating that motor skill learning induces a growing autonomy of sensorimotor systems accompanied by a decrease in the activation of cognitive control hubs \cite{Bassett:wx}. Early in learning, the visual and motor sub-networks are highly interconnected, and the connector hubs in cognitive control areas are highly active, potentially tuning and parsing connectivity between modules. Later in learning, the hubs are no longer required and the modules become disconnected and more autonomous. The faster this occurs, the faster the individual learns. Several recent studies across many different laboratories now provide additional evidence associating non-primary regions (especially but not solely in frontoparietal cortex) with both network reconfiguration and behavior on tasks demanding higher order cognitive function \cite{gerraty2018dynamic,braun2015dynamic,shine2016temporal,pedersen2018multilayer,alavash2015persistency}. The capacity for frontoparietal regions to enact this network-level control has been posited to stem from the specific pattern of white matter connections emanating from those regions to the rest of the brain \cite{gu2015controllability}. Specifically, using network control theory \cite{kim2018role,tang2018control}, the regions of the brain predicted by their pattern of white matter connections to most easily induce difficult state transitions in system function are located in frontoparietal areas. In further support of this hypothesis, individuals whose brains have greater network controllability (as calculated from the theory parameterized by their unique white matter connectivity) also have greater cognitive performance in general \cite{tang2017developmental} and cognitive control in particular \cite{cornblath2018sex,cui2018optimization}. Collectively, these studies support the notion that network control, instantiated on human white matter connectivity, provides a mechanistic explanation for cognitive control, and its associated influence on the activity and connectivity of other areas.

\section{Bridging Scales with Networks}

Finally, it is critical to note that networks form a single and natural mathematical language with which to frame questions within and across multiple scales of neural function. The benefit of framing mechanistic questions with this math is that the units involved are clearly specified, the edges between units within a scale are the channels along which work can be done, and the edges between scales allow the units in one scale to do work on the units in other scales. In other words, multi-scale networks provide a scaffold on which causal inter-scale dynamics can occur, allowing us to generate parsimonious multiscale descriptions and mechanistic explanations. 

\begin{figure}
\centering
\includegraphics[width=10cm]{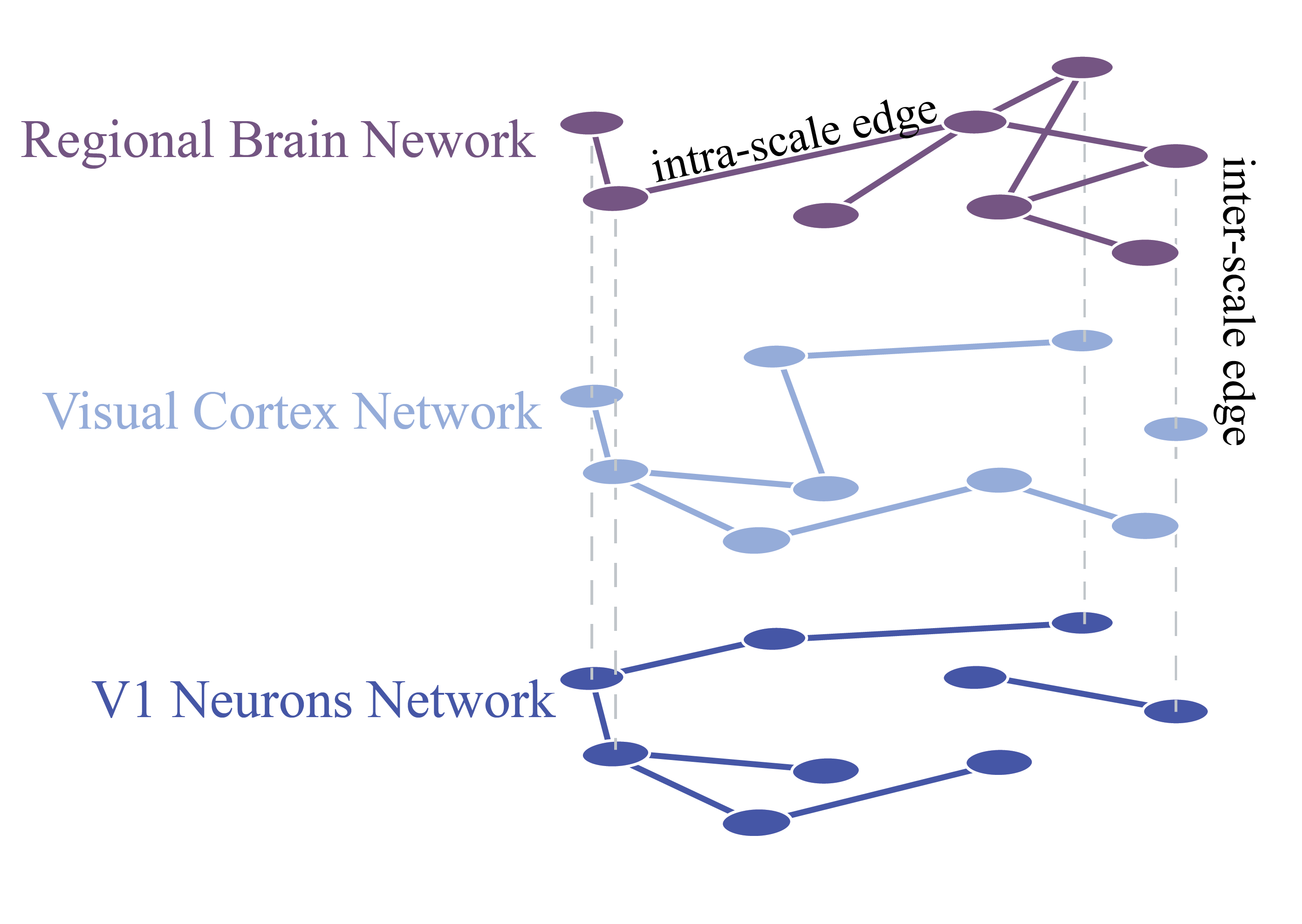}
\caption{\textbf{A multi-scale network model of relationships within and between scales} Multi-scale networks are a natural language in which to simultaneously model networks that exists at different scales in the brain. Here, edges within a scale indicate interactions between those two nodes within a scale, while edges between two scales indicate an interaction between those two nodes across scales. In this example, regional brain connectivity exists at the macro-scale, visual cortex connectivity at the meso-scale, and V1 neuronal connectivity at the micro-scale. Here, the connectivity of a node at the macro-scale could impact the connectivity of a node at the meso-scale, which could impact the connectivity of a node at the micro-scale.}
\end{figure}

Take vision as an example: one can explain much -- but not all -- of vision by what occurs in visual cortex. While artificial neural networks can reproduce some functions of cells in visual cortex \cite{kriegeskorte2015deep}, those functions also depend on the activity and function of other parts of cortex. For example, vision cannot be completely explained without also including a model of attentional inputs from frontal cortex. Yet, the computational language that we use to explain the cellular functions of vision (convolutional neural networks) is not the same computational language that we use to explain the regional functions of attention (top-down control and gating theories). The lack of a common language in which to frame explanations across scales and functions holds the field back; we can construct a list of such disjoint explanations, but at the end of the day they remain just that, a list. What we would instead like to have is a set of interdigitated explanations from which we can deduce the mechanisms by which scales and functions causally impact one another. 

Network science provides a common language with which to interdigitate explanations. By encoding the brain as a network, we can reason about vision processes in occipital cortex and attentional processes in frontal cortex using the same language. Moreover, we can reason about how the regional network underpinning attentional processes in frontal cortex can causally impact the cellular network underpinning visual processes in occipital cortex, largely because network science has specific tools for multi-layers networks that involve links between the layers. These inter-scale, inter-function connections in a multilayer network encoding of the brain can comprise the conduits along which work can be done. We can model how macro-scale network interactions, like connector hub tuning, influence micro-scale interactions, like neural tuning in V1, in a single model. If, however, the two phenomena were not translated into the same language (such as the language of network science), this knowledge would remain out of reach.

A notable secondary benefit of the shared language is that we can begin to deduce general principles of brain function shared across scales and functions. Perhaps cellular-level tuning functions in visual cortex share similar mechanisms with regional-level tuning functions that connector hubs may enact to control brain-wide connectivity. As we described earlier, we can use the language of network science to formalize the notion that connector hubs have the capacity to tune connectivity for integration in a modular network, and this notion provides a possible mechanism for the commonly studied process of top-down attentional control. We can speculate that neural tuning, both within visual cortex and across the cortex, is a general principle of brain function: primary regions tune for sensory, association, or motor information, while transmodal regions (here, connector hubs) tune for connectivity patterns that allow for that information to be integrated across modular processors. By articulating explanations across scales in the same network language, we can begin to test such speculations with the goal of discovering general principles of brain function that exist across scales, and distinguishing them from principles that exist at only a single scale. 

The reasonableness and biological plausibility of interdigitated explanations are particularly salient when one considers evolution. The processes of natural selection did not drive the evolution of single regions independently, but instead led to the formation of the entire brain simultaneously. While visual cortex exists so that organisms can see, the brain exists so that organisms can create offspring who have a high probability of reproducing themselves. Moreover, visual cortex develops alongside and dependent on cascades of neurodevelopmental processes that span the entire brain. In other words, both visual cortex and other areas of cortex experience some of the same evolutionary pressures, impacting cellular and regional scales, which could drive similar patterns of connectivity across scales and across regions. The notion that shared causes can drive shared patterns of connectivity can also provide insight as we consider neural systems across species. As described earlier, modular connectivity patterns with a diverse club of tightly interlinked connector hubs have been identified in the cellular network of neurons in \emph{C. elegans} as well as the regional network of areas in the human brain; it may be that this architecture is nature's solution to integration in a modular network. 

Of course it is important to admit that network science is not the only mathematical language with which to describe the brain. Yet, network science has marked advantages over other options in part because of its authenticity; no metaphors are needed to link the physical organ of the brain to the mathematics of network science. The brain is a network, across species, and across scales. But perhaps it is worth also acknowledging that brains are extremely complicated networks. It requires a formal theoretical apparatus like network science to represent the brain in a way that is intelligible to us and in a way that allows us to link network features to one another and to behavior, compare brains across species, and simulate the evolution of networks to better understand the reasons for their architecture.

\section{Conclusion}

In conclusion, the strength of network neuroscience is that it can take complex networks and reduce this complexity by describing the network succinctly. Network concepts help us to turn the messy reality of the brain into quantitative variables that make the search for correlations that confirm mechanistic models tractable. Correlational analyses in network neuroscience can provide evidence in support of causal mechanisms, particularly when combined with analyses that demonstrate necessity. Efforts to test mechanistic models via diverse types of analyses can provide diverse types of evidence. Critically, because the mechanistic explanations in human network neuroscience are framed in a language that we can also use to study how neurons work at the smaller scales of visual cortex or simpler organisms, it is possible to obtain general principles of brain function that are true across scales.

More generally, it is of fundamental importance to understand and articulate the nature of explanations that are accessible to distinct areas of science and their associated experimental, computational, or theoretical approaches. Here we have attempted to clarify the distinctions between an explanation and the evidence supporting that explanation. Moreover, we have sought to distinguish between a mechanism and the scale at which that mechanism exists. Drawing on extensive work in the field of philosophy, we have framed our discussion largely within the context of emerging approaches from network science that are proving particularly interesting and satisfying for many neuroscientists \cite{bassett2018nature}. In the future, we envision increasing clarity on the network mechanisms that are pertinent to brain function at large scales, intermediate scales, and small scales, and a broadly held positive valuation of mechanisms irrespective of scale. We also envision increasing clarity on how mechanisms at one scale interdigitate with mechanisms at the scale above and the scale below, fostered by network analyses that formalize the scales in the same language and provide a language to link scales to one another. Another way we see neuroscience progressing in the coming years is that our macro-scale findings can guide micro-scale analyses that involve necessitative evidence or manipulations. Finally, we hope that this work serves as an example of how precise language and distinctions from the philosophy of science can be combined with recent advances in neuroscience to propel the field forward.

\bibliography{main}



\end{document}